%% file: main.tex
\documentclass[%
 aip,
 amsmath,amssymb,
 reprint,%
]{revtex4-1}

\usepackage{graphicx}
\usepackage{dcolumn}
\usepackage{bm}

\usepackage{mathtools}
\usepackage[utf8]{inputenc}
\usepackage[T1]{fontenc}
\usepackage{mathptmx}
\usepackage{etoolbox}
\usepackage{xcolor}
\usepackage{graphics}
\usepackage{siunitx}
\usepackage{dcolumn}

\makeatletter
\def\@email#1#2{%
 \endgroup
 \patchcmd{\titleblock@produce}
  {\frontmatter@RRAPformat}
  {\frontmatter@RRAPformat{\produce@RRAP{*#1\href{mailto:#2}{#2}}}\frontmatter@RRAPformat}
  {}{}
}%
\makeatother

\newcommand{\vecr}{\bm{r}} 

\begin{document}

\title[Evolutionary Integral Equations]{Evolutionary optimization of the Verlet closure relation for the hard-sphere and square-well fluids}

\author{Edwin Bedolla}
 \email{ea.bedollamontiel@ugto.mx}
\affiliation{ 
División de Ciencias e Ingenierías, Universidad de Guanajuato, Loma del Bosque 103, 37150 León, Mexico
}%
\author{Luis Carlos Padierna}%
\affiliation{ 
Departamento de Ingeniería Química, Electrónica y Biomédica, División de Ciencias e Ingenierías, Universidad de Guanajuato, Loma del Bosque 103, 37150 León, Mexico
}%

\author{Ramón Castañeda-Priego}
\affiliation{%
Departamento de Ingeniería Física, División de Ciencias e Ingenierías, Universidad de Guanajuato, Loma del Bosque 103, 37150 León, México
}%

\date{\today}

\begin{abstract}
    The Ornstein-Zernike equation is solved for the hard-sphere and square-well fluids using a diverse selection of closure relations; the attraction range of the square-well is chosen to be $\lambda=1.5.$
    In particular, for both fluids we mainly focus on the solution based on a three-parameter version of the Verlet closure relation [Mol. Phys. \textbf{42}, 1291–1302 (1981)]. To find the free parameters of the latter, an unconstrained optimization problem is defined as a condition of thermodynamic consistency based on the compressibility and solved using Evolutionary Algorithms. For the hard-sphere fluid, the results show good agreement when
    compared with mean-field equations of state and accurate computer simulation results; at high densities, i.e., close to the freezing transition, expected (small) deviations are seen. In the case of the square-well fluid, a
    good agreement is observed at low and high densities when compared with event-driven molecular dynamics computer simulations. For intermediate densities, the explored closure relations vary in terms of accuracy. Our findings suggest that a modification of the optimization problem to include, for example, additional thermodynamic consistency criteria could improve the results for the type of fluids here explored.
\end{abstract}

\maketitle

\input{files/introduction.tex}

\input{files/theory.tex}
\input{files/evolutionary.tex}

\input{files/methodology.tex}

\input{files/results.tex}

\section{Concluding remarks and outlook}\label{sec:conclusions}
In this work, the three-parameter Verlet closure relation (Eq.~\eqref{eq:three-mv-abs}) was used to study the equation of state of the hard-sphere fluid, as well as for the square-well fluid with a well range $\lambda=1.5$ at a reduced temperature $T^{*}=1.5$ (well above the critical temperature). To fix the parameters in the closure relation, Evolutionary Algorithms were used to solve a bounded unconstrained optimization problem
by enforcing thermodynamic consistency via the inverse isothermal compressibility. For the hard-sphere fluid, it was shown that the closure relation reproduces the
original results by Verlet~\cite{verletIntegralEquationsClassical1981}, which is the
closure relation that gave good results when compared with computer simulations and mean-field
equations of state, at least from low to intermediate densities. For the square-well
fluid, results obtained with the Zerah-Hansen and the three-parameter Verlet closure relation
were the more accurate when compared with computer simulations. At intermediate densities, both closure relations along with the Mean Spherical Approximation seemed to oscillate around the simulation results and provided good results.

Still, for a more conclusive study, several well depths and attractive ranges should be studied and at different thermodynamic state points. It should be possible to study these systems, as it has been shown in this work that the methodology works for different short range fluids.
This methodology can be readily extended to other kinds of intermolecular potentials, such as long-ranged potentials like the Lennard-Jones fluid. The advantage of the proposal presented here is that by defining a general unconstrained optimization problem, and with the aid of DE algorithms, the same methodology should work, regardless the number of parameters and the specific mathematical form of the intermolecular potential.

In both fluids here considered, it seemed that a combination of closure relations might be beneficial for an
accurate estimation of the compressibility factor. Similarly, the three-parameter Verlet closure relation should be included.
The mixing of the two and three parameter versions of the Verlet closure relation
could lead to improved results for the hard-sphere fluid. For the square-well fluid, the combination of the Zerah-Hansen and the three-parameter version could be used to get better estimations.
The fact that Evolutionary Algorithms are used to solve the optimization problems implies that there is little
overhead when including more parameters, as these algorithms are perfectly suited for
high-dimensional and nonlinear optimization problems. We believe that the success of the
three-parameter Verlet closure relation is closely related to the diagram expansion and
that it has a better suited functional form to estimate the compressibility factor in both
classical fluids.

Last, but not least, we should stress out that although this is not the first attempt to combine both computational intelligence and the integral equations theory formalism, this contribution opens up a new venue to use robust optimization techniques to solve the Ornstein-Zernike equation and to build a more general scheme that searches physical solutions that guarantee thermodynamic consistency. As the choice of optimization algorithm is an important part of this framework, a new route to explore would be a direct comparison between the most common Evolutionary Computation algorithms; work in this direction is currently in progress.

\begin{acknowledgements}
R. C.-P. acknowledges the financial support provided by the Consejo
Nacional de Ciencia y Tecnolog{\'i}a (CONACYT, M{\'e}xico) through
Grants Nos. 237425 and 287067.
\end{acknowledgements}

\bibliography{bibliography.bib}

\end{document}

%% file: files/introduction.tex
\section{Introduction}\label{sec:introduction}
Integral equation theories have been a recurring theoretical framework to study the structural and thermodynamic properties of both simple and complex fluids \cite{hansenTheorySimpleLiquids2013}. The most common framework is the Ornstein-Zernike (OZ) equation~\cite{hansenTheorySimpleLiquids2013}.
This is in part due to its theoretical formulation, and modern techniques for providing an efficient numerical solution~\cite{gillanNewMethodSolving1979,zerahEfficientNewtonMethod1985,labikRapidlyConvergentMethod1985,kelleyFastSolverOrnstein2004}. However, the OZ equation must be paired with a closure relation to get a physical solution~\cite{hansenTheorySimpleLiquids2013}. The use of a closure relation relies on approximating some physical information that produces results that commonly depend on the physical or mathematical path used to obtain a solution, leading in some cases to inconsistencies. This is normally referred to as thermodynamic inconsistency~\cite{hansenTheorySimpleLiquids2013}, and it constitutes one of the biggest challenges to the use of integral equation theories.

To solve the problem of thermodynamic inconsistency, several closure relations have been developed, most of the time tailored to specific intermolecular fluids~\cite{ladoSolutionsReferencehypernettedchainEquation1983,rogersNewThermodynamicallyConsistent1984b,zerahSelfConsistentIntegral1986,lombaOrnsteinZernikeEquationsSimulation1993}. One such case is the hard-sphere fluid, which serves as an invaluable reference model for understanding simple and complex fluids~\cite{hansenTheorySimpleLiquids2013,santosStructuralThermodynamicProperties2020a}. The Percus-Yevick~\cite{percusyevick1958} approximation is one of the first closure relations used within the OZ formalism to provide an analytical solution for the hard-sphere fluid.
Still, even though the solution is exact, the problem of thermodynamic inconsistency is present when it is used to determine, for example, either the pressure or compressibility~\cite{hansenTheorySimpleLiquids2013}. This means that the latter computed through different routes might give different results, which is not desired. The main approach to solving this issue is to enforce thermodynamic consistency using different routes for all possible state variables. Verlet proposed using the first exact virial coefficients to provide a closure for the hard-sphere fluid, creating a semiphenomelogical approximation that works quite well, as shown in the original work~\cite{verletIntegralEquationsClassical1980}, where the approximation was compared directly with computer simulations.
A different approach is to drive consistency by obtaining the same value for thermodynamic pressure using multiple routes, which has been extensively explored before~\cite{verletIntegralEquationsClassical1981,verletIntegralEquationsClassical1982,martynovExactEquationsTheory1983,rogersNewThermodynamicallyConsistent1984b,balloneAdditiveNonadditiveHard1986,martynovDifferentialConditionThermodynamic1993,duhIntegralEquationTheory1995,leeAccurateIntegralEquation1995}. This is the so-called partial thermodynamic consistency, which is simple to implement and use, but it has the disadvantage that other state variables might still give different results depending on the route used to calculate them~\cite{mcquarrieStatisticalMechanics2000,hansenTheorySimpleLiquids2013}.
For more robust consistency requirements, free energy must also yield the same result through different routes, and other proposals that include this information have been studied for the case of the hard-sphere fluid~\cite{moritaTheoryClassicalFluids1958,martynovDifferentialConditionThermodynamic1993,tsedneeClosureOrnsteinZernikeEquation2019}.

Most of these proposals deal with the hard-sphere fluid, and the closure relation must be chosen for each particular intermolecular potential. The work by~\citeauthor{a.goodallDatadrivenApproximationsBridge2021}~\cite{a.goodallDatadrivenApproximationsBridge2021} is a recent attempt to work towards a unified framework of closure relations, using the formalism and modern tools of Machine Learning (ML).
By collecting a large data set of different interactions given by a diverse range of attractive and repulsive intermolecular potentials, a closure relation can be obtained that could potentially work in most cases by fitting a neural network to the training data.
Yet, the amount of computation needed to perform such tasks is too large if a single intermolecular potential needs to be studied. Similarly, thermodynamic consistency is not enforced in any way, only relying on the results obtained from computer simulation data. Still, the approach is a novel way using data-driven methods in the theory of simple liquids, paving the way to explore other kinds of ML methods that could be used to gain a better understanding of the role of bridge functions and closure relations.

In this work, we follow the original proposal by Verlet~\cite{verletIntegralEquationsClassical1981}, and rely on the particular formulation of the modified Verlet (MV) approximation, which has proven to be an excellent approximation in several applications of the hard-sphere fluid~\cite{lopez-sanchezDemixingTransitionStructure2013a,zhouLocalStructureThermodynamics2019a,perera-burgosAssessmentMicrostructureDepletion2016a,zhouIntegralEquationTheory2020a,meiThermodynamicsStructureDynamics2020}. Through the enforcement of compressibility consistency, the free parameters of the closure relation are found. To find these parameters, an unconstrained optimization formulation of the problem is defined and solved using Evolutionary Computation algorithms~\cite{fogelWhatEvolutionaryComputation2000}. It is shown that the case of a monocomponent hard-sphere fluid is solved correctly, finding results consistent with those reported by Verlet, i.e., the closure relation works well with small and expected deviations at high densities (close to the freezing transition). Then, the same closure relation is used with the square-well fluid for a particular attraction range parameter, namely, $\lambda=1.50$. The closure relation is found to give good results when compared with event-driven molecular dynamics computer simulations and other closure relations.

This work is organized as follows. In Sec.~\ref{sec:int-equations}, the theoretical framework of integral equations of liquids is developed, as well as the closure relations and the role of the bridge function. In Sec.~\ref{sec:evo-opt}, Evolutionary Optimization is introduced along with the methods used to solve global unconstrained optimization problems. In Sec.~\ref{sec:methods}, the numerical methods for solving the OZ equation are presented, as well as the computer simulation methods and the details to obtain the results presented in this paper. The results and discussions for the hard-sphere and square-well fluids are provided in Sec.~\ref{sec:results}. The work is closed with remarks and outlooks for future work in Sec.~\ref{sec:conclusions}.

%% file: files/theory.tex
\section{Integral equations theory}\label{sec:int-equations}
\subsection{Ornstein-Zernike equation}

The OZ equation provides a straightforward route to calculate the static properties of a liquid in equilibrium, using both direct and indirect contributions of the interacting particles. We briefly describe the theoretical formalism here, but there are excellent descriptions of it in standard textbooks~\cite{mcquarrieStatisticalMechanics2000,hansenTheorySimpleLiquids2013}.
For an isotropic, monocomponent closed system at temperature $T$, the OZ equation reads as~\cite{hansenTheorySimpleLiquids2013},
\begin{equation}
    h(r)=c(r)+\rho \int c\left(\lvert \vecr - \vecr^{\prime} \rvert \right) \, h(\vecr^{\prime}) \, d \vecr^{\prime} \; ,
    \label{eq:oz-eq}
\end{equation}
where $\rho=N/V$ is the number density, with $N$ being the number of particles contained in a volume $V$.
The OZ equation in Eq.~\eqref{eq:oz-eq} introduces the total correlation function, $h(r)$, and the direct correlation function, $c(r)$.
With these functions, the indirect correlation is defined as \(\gamma(r) \equiv h(r)-c(r)\), as well as the radial distribution function (RDF), defined as \(g(r)=h(r)+1\).

To provide a solution to Eq.~\eqref{eq:oz-eq}, a second expression must be introduced that relates the correlation functions to the pairwise intermolecular potential, $u(r)$. This new expression is the so-called closure relation, and in its most general form is defined as~\cite{hansenTheorySimpleLiquids2013},
\begin{equation}
    c(r)=\exp{\left[- \beta u(r) + B(r) + \gamma(r)\right]}-\gamma(r)-1 \; ,
    \label{eq:exact-closure}
\end{equation}
where $B(r)$ is known as the \emph{bridge function}, \(\beta=1/k_{B} T\),
with $k_{B}$ the Boltzmann constant. When attempting to solve the OZ equation, an approximation for $B(r)$ is required~\cite{hansenTheorySimpleLiquids2013}. This is due to the fact that the bridge function can only be expressed exactly as an infinite power series in density~\cite{hansenTheorySimpleLiquids2013}. The most common closure relations are the so-called Hypernetted Chain (HNC),
\begin{equation}
    B_{HNC}(r)=0 \; ,
    \label{eq:hnc}
\end{equation}
and the Percus-Yevick (PY) equation~\cite{percusyevick1958},
\begin{equation}
    B_{PY}(r)=\ln{\left[ \gamma(r) + 1 \right]} - \gamma(r) \; .
    \label{eq:percus-yevick}
\end{equation}
The latter can also be obtained by assuming that $B(r)=0$ in Eq.~\eqref{eq:exact-closure} and linearizing the resulting argument of the exponential \cite{perera-burgosAssessmentMicrostructureDepletion2016a}.

\subsection{Thermodynamic consistency}
When approximations for the bridge function are used together with the OZ equation, most of 
the time the solution shows a phenomenon where, if a thermodynamic observable is computed through two or more \emph{routes}, the results may vary drastically. This is not the desired behavior of the closure relation. Rather, it is expected that the closure relations 
could, in fact, provide accurate results regardless the thermodynamic route they are computed with.

For example, a standard route for computing the thermodynamic pressure, $P$, is the so-called \emph{virial equation}~\cite{hansenTheorySimpleLiquids2013}
\begin{equation}
    \beta P = \rho - \frac{\rho}{6} \int_{0}^{\infty} \frac{\partial \beta u(r)}{\partial r} \, g(r) \, r \, d \vecr \, .
    \label{eq:pressure-gr}
\end{equation}
A related thermodynamic quantity, the \emph{isothermal compressibility}, is defined as~\cite{kondepudi2014modern},
\begin{equation}
    \chi_{T} = - \, \frac{1}{V} { \left( \frac{\partial V}{\partial P} \right) }_{T} =
    \frac{1}{\rho} { \left( \frac{\partial \rho}{\partial P} \right) }_{T}
    \; .
    \label{eq:chi-thermo}
\end{equation}
Following Eq.~\eqref{eq:chi-thermo}, together with Eq.~\eqref{eq:pressure-gr}, $\chi_{T}$ can be computed through these definitions. We refer to this way of computing $\chi_{T}$ as the \emph{pressure route}.

A different way of computing $\chi_{T}$ is by using correlation functions~\cite{hansenTheorySimpleLiquids2013},
\begin{equation}
    \frac{\beta}{\left( \chi_{T} \, \rho \right)} = 1 - \rho \int d \vecr \, c(r)
    \; ,
    \label{eq:compress-gr}
\end{equation}
where $c(r)$ is the direct correlation function introduced in the OZ equation, Eq.~\eqref{eq:oz-eq}.
Using Eq.~\eqref{eq:compress-gr} to compute $\chi_{T}$ is called the \emph{compressibility route}.

Standard closure relations, such as HNC or PY, lead, in general, to different results for the hard-sphere fluid when using the compressibility and pressure routes~\cite{hansenTheorySimpleLiquids2013}.
When both routes produce the same quantity for the isothermal compressibility for a given interaction potential, it is said that the OZ equation along with the closure relation used is partially thermodynamically consistent. 

\subsection{Modified Verlet bridge function}
Between 1980 and 1981, Loup Verlet published a series of two papers dedicated to the integral equations of hard-sphere fluids and introduced what is now called the modified Verlet (MV) bridge 
function~\cite{verletIntegralEquationsClassical1980,verletIntegralEquationsClassical1981}.
The original proposition for the bridge function by Verlet is~\cite{verletIntegralEquationsClassical1980},
\begin{equation}
    B(r) = - \frac{A \, {\left[\gamma(r)\right]}^{2} \, / 2}{1 + B \, \gamma(r) \, / 2}
    \; .
    \label{eq:verlet-params}
\end{equation}
By fitting the exact values of the first five virial coefficients for the hard-sphere fluid, Verlet reached the conclusion that 
the values for \(A\) and \(B\) that would reproduce the expected results up to the fluid-solid transition 
were the values \(A = 1 \, , \, B = 4/5\), which turn the original closure relation in Eq.~\eqref{eq:verlet-params} into,

\begin{equation}
    B(r) = - \frac{0.5 \, {\left[\gamma(r)\right]}^{2}}{1 + 0.8 \, \gamma(r)}
    \; .
    \label{eq:mVerlet}
\end{equation}

This turned out to be a highly accurate bridge function approximation for the 
hard-sphere fluid, as shown by Verlet when comparing the results obtained with equations
of state and computer simulations~\cite{verletIntegralEquationsClassical1980}.
Some of the advantages of this approximation are the fact that it is simple, efficient, and 
there are no free parameters to be fixed, contrary to the case of, for example, the
Rogers-Young~\cite{rogersNewThermodynamicallyConsistent1984b} and 
Zerah-Hansen~\cite{zerahSelfConsistentIntegral1986} closure relations.
It has been shown that even when hard-sphere mixtures are 
considered~\cite{lopez-sanchezDemixingTransitionStructure2013a,zhouLocalStructureThermodynamics2019a}, the MV bridge function produces excellent results. 
Similar results have been obtained for the case of hard-disk fluids and their mixtures~\cite{perera-burgosAssessmentMicrostructureDepletion2016a}.
More recently, the MV closure relation has been used in overcompressed and supercooled metastable hard-sphere fluids~\cite{zhouIntegralEquationTheory2020a}, as well as in the study of correleations and nonuniversal effects in glass-forming liquids~\cite{meiThermodynamicsStructureDynamics2020}, with good results overall.

The problem lies, as with most closure relations, in the fact that the bridge function in Eq.~\eqref{eq:mVerlet} is not thermodynamically consistent. There are other closure relations 
which are already thermodynamically consistent, e.g., the reference Hypernetted 
Chain approximation~\cite{ladoSolutionsReferencehypernettedchainEquation1983}. This 
approximation comes from first principles, and the thermodynamic consistency comes from the 
fact that the reference potential is chosen such that the free energy of the fluid is 
minimized~\cite{ladoSolutionsReferencehypernettedchainEquation1983}. Still, the problem with this closure relation, although accurate, is that it 
might be hard to implement and use in several different scenarios, such as being useful for different kinds of intermolecular potentials. Also, the use of a reference potential means that a approximation for the bridge function must be used for that reference potential. In the original work by~\citeauthor{ladoSolutionsReferencehypernettedchainEquation1983}~\cite{ladoSolutionsReferencehypernettedchainEquation1983}, the hard-sphere fluid was used as the reference potential, but any other reference potential can be used as long as there is an accurate bridge function for that fluid.

In the second paper in the series~\cite{verletIntegralEquationsClassical1981}, Verlet developed a second approach to the MV closure relation. The purpose of the reformulation was to provide a closure relation on first principles, and make the approximation partially thermodynamically consistent by introducing free parameters. The basic idea is to use Padé approximants for the bridge function, then the coefficients of such polynomials will be obtained through a minimization process that successfully reproduces the virial and compressibility routes. For more details on the formulation, we refer the reader to the original work by Verlet~\cite{verletIntegralEquationsClassical1981}.

The form of the improved closure relation is a generalized three-parameter version of the original closure relation described by Eq.~\eqref{eq:mVerlet},
\begin{equation}
    B(r) = A \, {\gamma(r)}^{2} \, \left[\frac{1 + B \, \gamma(r)}{1 + C \, \gamma(r)} \right]
    \; .
    \label{eq:three-mv}
\end{equation}

To obtain the values for $A,B,C$, Verlet used a series of steps to approximate the value of the thermodynamic pressure. In essence, the idea is to fix a value for $A$ and $B$, and then vary $C$. This process is then performed once again to find $B$, and finally to determine $A$. To obtain an accurate estimation of the equation of state of the hard-sphere fluid, an intermediate step based on using the Carnahan-Starling equation of state~\cite{mcquarrieStatisticalMechanics2000} was needed. The results obtained from this procedure revealed that good agreement with previous results reported were achieved, although the compressibility factor is underestimated at high densities.

\subsection{The Kinoshita variation}
The Kinoshita variation to the MV closure relation reads~\cite{kinoshitaInteractionSurfacesSolvophobicity2003},
\begin{equation}
    B(r) = - \frac{0.5 \, {\left[\gamma(r)\right]}^{2}}{1 + 0.8 \, \left\lvert \gamma(r) \right\rvert}
    \; ,
    \label{eq:kinoshita-eq}
\end{equation}
and introduces the absolute value \(\left\lvert \cdot \right\rvert\) in the denominator, which increases the numerical stability of the closure relation, avoiding the possibility that the denominator becomes zero.

It is important to note that even though this approach seems to be a good approximation for
the particular case of the hard-sphere fluid, in general, a virial expansion based on the
evaluation of the virial coefficients does not lead to a proper equation of state. The 
reason for this is that the virial expansion is an infinite series, but computationally 
only truncated series can be dealt with, and so the series has to be truncated somewhere,
dropping accuracy and physical relevance for higher order terms in the expansion, which are important at high particle concentrations.

%% file: files/evolutionary.tex
\section{Evolutionary optimization}\label{sec:evo-opt}
The idea behind Evolutionary Computation (EC) is to use the powerful mechanism of evolution 
and apply it to complex problems, using modern computational and mathematical 
resources to facilitate and empower a simplified version of evolution.~\cite{fogelWhatEvolutionaryComputation2000} In the most general 
case, evolution is roughly a two-step process: first comes \emph{variation} and then comes 
\emph{selection}~\cite{kacprzykSpringerHandbookComputational2015}. Algorithms based on EC use these properties, and, depending on how these 
variations and selections occur, the algorithms are named and used differently. In 
particular, for this work, the focus will be on \emph{evolution strategies} and their direct application to global optimization problems, which will be 
addressed next. However, there are also \emph{genetic algorithms}, 
\emph{evolutionary algorithms}, and others~\cite{kacprzykSpringerHandbookComputational2015}.

\subsection{Evolution Strategies}
The principal characteristics of Evolution Strategies (ES) are 
the ease of dealing with high-dimensional optimization problems and the low number of function 
evaluations. Also, unlike other EC algorithms, ES methods evaluate the fitness of the 
functions in batches instead of individually, allowing for more efficient computations, as well as a more efficient search of
solution space. The most prominent algorithm in the ES family of methods 
is the CMA-ES, or Covariance Matrix Adaptation-Evolution 
Strategy~\cite{hansenCMAEvolutionStrategy2006}.
In general, ES methods do two main things. First, to perform \emph{variation}, multivariate normal 
random vectors are used. Second, to carry out \emph{mutation}, these algorithms modify different 
aspects of that particular probability distribution, such as \emph{mean} and \emph{covariance}. One of the main disadvantages of the CMA-ES is its high computational cost, having to perform the inverse of the covariance matrix at each step of the solution.

\subsection{Differential Evolution}
\begin{figure}
    \centering
    \includegraphics[scale=0.4]{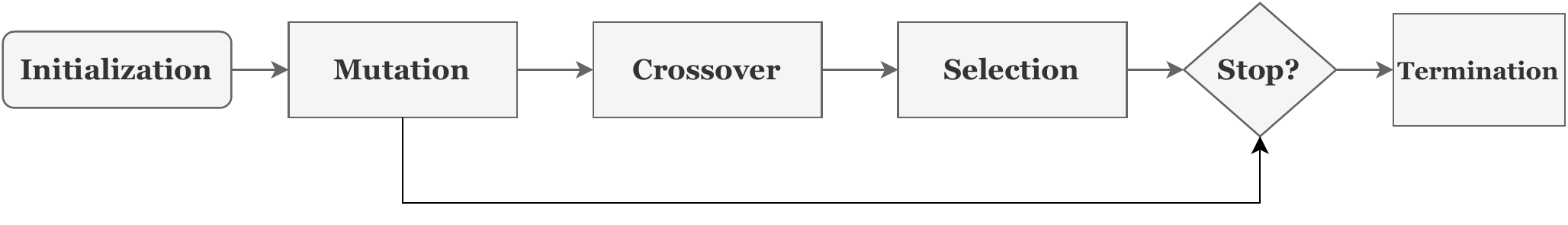}
    \caption{Flowchart of a generic DE optimization method.}
    \label{fig:de-flowchart}
\end{figure}

Differential Evolution (DE)~\cite{stornDifferentialEvolutionSimple1997,bilalDifferentialEvolutionReview2020} methods have been widely used as stochastic optimization methods due to their simplicity and efficiency in finding a solution in a short period of time. These methods perform the same steps as all ES methods, shown in Fig.~\ref{fig:de-flowchart}.

During \emph{initialization} a set of \(NP \in \mathbb{Z}\) candidate solutions is chosen at random, based on the upper and lower bounds of the search space. These solutions, also known as \emph{agents}, will be used throughout the complete optimization procedure.

Then comes the \emph{evolution} of the agents, starting with \emph{mutation}. During the mutation step, a \emph{donor} is computed at generation \(G\), as a weighted difference between two randomly chosen agents,
\begin{equation}
    V_{j}^{G} = X_{r1}^{G} + F \cdot \left( X_{r2}^{G}-X_{r3}^{G} \right) \; ,
    \label{eq:donor}
\end{equation}
where \(F \in [0,1]\) is a scaling vector, most commonly chosen to be \(F=0.8\)~\cite{bilalDifferentialEvolutionReview2020,ahmadDifferentialEvolutionRecent2022}. Labels \(r1, r2, r3 \in \{1,2,\cdots,NP \}\) represent different agents chosen at random from the set of available candidates. The index \(j\) takes the values \(1,2,\cdots,NP.\)

The next step in the evolution process is that of the \emph{crossover}. In this step, a new trial agent is created from the donor and a target agent. By selecting a crossover probability \(Cr \in [0,1]\), the new trial agent is generated as,
\begin{equation}
    U_{j}^{G} = \begin{dcases}
    V_{j}^{G} & rand \leq Cr \\
    X_{j}^{G} & rand > Cr \; ,
    \end{dcases}
    \label{eq:crossover}
\end{equation}
where \(rand\) is random number from a uniform probability distribution. This means that if the condition is met, the new trial agent is chosen instead of the agent already in the population.

The final step is called \emph{selection}, and it is based on evaluating the fitness of the function and the trial agent obtained with Eq.~\eqref{eq:crossover}. If the evaluation of the fitness is lower for the trial agent, then the trial agent \(U_{j}^{G}\) is kept and \(X_{j}^{G}\) is discarded. If not, the original target agent is kept.
This process is repeated for as many generations as needed, or when a given stopping criterion is met. In the area of numerical optimization, the most common choices for stopping criteria are running time of the algorithm; target fitness of the function evaluation; or number of function evaluations~\cite{nocedalNumericalOptimization2006}.

In most real applications, the simplicity of DE methods is its biggest disadvantage, and several enhancements have been researched and implemented since its invention, mostly focused on the crossover operation, which is critical to the performance of the DE methods~\cite{neriRecentAdvancesDifferential2010,dasDifferentialEvolutionSurvey2011a,dasRecentAdvancesDifferential2016}. In this work, a variation of the DE method using CMA-ES covariance matrix learning techniques~\cite{wangDifferentialEvolutionBased2014} is used to solve the unconstrained global optimization problems. The implementation used in this work is the radius limited DE method from the software BlackBoxOptim.jl~\cite{bbopt}.

%% file: files/methodology.tex
\section{Methodology}\label{sec:methods}
By enforcing that the isothermal compressibility 
computed from Eq.~\eqref{eq:chi-thermo}, with the pressure calculated with Eq.~\eqref{eq:pressure-gr}, and the isothermal compressibility given by Eq.~\eqref{eq:compress-gr} are equal to 
some arbitrary precision, closure relations can be partially thermodynamically consistent. The standard way of alleviating the problem of thermodynamic 
inconsistency is to compute $\chi_{T}$ using both routes and adjust some free (non-physical) parameter, or function, 
such that both thermodynamic routes yield the same result. In other words, both results must come from the same free energy functional~\cite{hansenTheorySimpleLiquids2013}. This is normally achieved by enforcing not only one, but several thermodynamic routes.

In this contribution, the proposal is to use the three-parameter variation of the Verlet closure approximation, Eq.~\eqref{eq:three-mv},
\begin{equation}
    B(r) = A \, {\gamma(r)}^{2} \, \left[\frac{1 + B \, \gamma(r)}{1 + C \, \left\lvert \gamma(r) \right\rvert} \right]
    \; ,
    \label{eq:three-mv-abs}
\end{equation}
where we have kept the absolute value in the denominator to ensure numerical stability. As this closure relation is an improvement to the original two-parameter version, this seems like an ideal candidate for the EC optimization method. Also, following the results from Verlet~\cite{verletIntegralEquationsClassical1980,verletIntegralEquationsClassical1981}, this closure relation should yield good results not only for the hard-sphere fluid. To the best of our knowledge, this closure relation has not been implemented for the square-well fluid yet.

To find the new free parameters \(A, B, C,\) partial thermodynamic consistency will be imposed through the use of the pressure and compressibility routes in order to obtain the same value for the isothermal compressibility.
In the special case when \(A=1/2, \, B=0, \, C=4/5,\) we recover the original two-parameter version of the Kinoshita variation, Eq.~\eqref{eq:kinoshita-eq}.

\subsection{Optimization setup}
\begin{figure*}
    \centering
    \includegraphics[width=\linewidth]{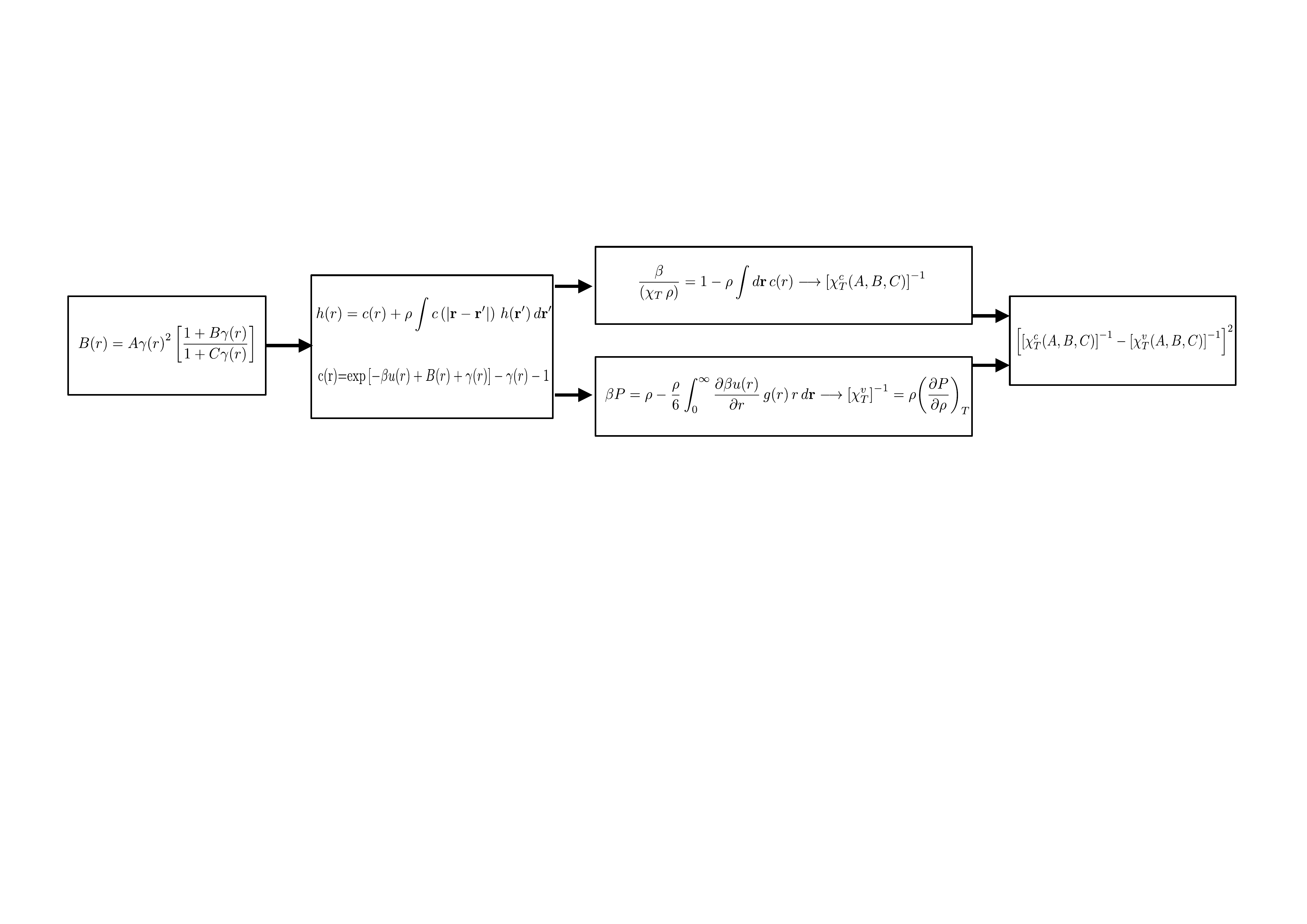}
    \vspace{-6.5cm}
    \caption{Schematic of the black-box function used for the optimization procedure implemented to solve the OZ equation, Eq.~\eqref{eq:oz-eq}, together with the three-parameter Verlet closure relation, Eq.~\eqref{eq:three-mv-abs}. The procedure is as follows. First, the three parameters $A,B,C$ are chosen randomly from a defined search space. Then, the closure relation is fixed and the OZ equation is solved. Using the compressibility and virial routes, the inverse isothermal compressibility is calculated. Finally, the squared difference between both routes is calculated and this is the value to be minimized, as explained in the text.}
    \label{fig:black-box-function}
\end{figure*}

In order to solve the problem of finding the free parameters \(A, B, C,\) in Eq.~\eqref{eq:three-mv-abs} that would provide
partial thermodynamic consistency, a special type of setup needs to be implemented.
It has already been stated that we seek the compressibility and virial routes to give the same result. This can be
formulated as an unconstrained optimization problem, if a particular physical observable is chosen.

Let ${\left[\chi_{T} (\bm{\alpha})\right]}^{-1}$ be the inverse isothermal compressibility, and
$\bm{\alpha}=\{A,B,C\} \in \mathbb{R}^3$ is the vector of free parameters, also known as the design vector.
A loss function can be defined as the squared difference between both quantities as follows,
\begin{equation}
    f(\bm{\alpha}) \equiv {\left[
        {\left[\chi_{T}^{c} (\bm{\alpha})\right]}^{-1} - {\left[\chi_{T}^{v} (\bm{\alpha})\right]}^{-1} \right]}^2
    \; .
    \label{eq:loss-function}
\end{equation}
where ${\left[\chi_{T}^{c} (\bm{\alpha})\right]}^{-1}$ is the inverse isothermal compressibility computed through the
compressibility route, while ${\left[\chi_{T}^{v} (\bm{\alpha})\right]}^{-1}$ is computed through the virial route.
Then, an unconstrained optimization problem that minimizes the function $f(\bm{\alpha})$ can be simply defined as,
\begin{equation}
    \underset{\bm{\alpha}}{\text{minimize}} \; f(\bm{\alpha})
    \; ,
    \label{eq:optimiziation-chi}
\end{equation}
In other words, for a fixed set of parameters $\bm{\alpha}$, the inverse isothermal compressibility is computed
via both routes, and its difference squared is calculated. The goal of the optimization problem is to minimize this
difference. If a good set of parameters is found, then the difference will be ideally zero or close to zero, which
in turn would mean that both routes are giving the same result, up to an arbitrary numerical tolerance.

Now, to solve the optimization problem, the numerical procedure needs to be included in a single
function evaluation. To achieve this, a black-box function is introduced that performs the following steps:
\begin{enumerate}
    \item Take as input three fixed values, $A,B,C,$ and use them to define the closure relation in Eq.~\eqref{eq:three-mv-abs}.
    \item Solve the OZ equation with the closure relation provided in the previous step. This will output two quantities, the radial distribution function, \(g(r)\), and the direct correlation function, \(c(r)\).
    \item Using Eq.~\eqref{eq:pressure-gr} and Eq.~\eqref{eq:chi-thermo}, ${\left(\chi_{T}^{v}\right)}^{-1}$ is computed.
    \item Then, using Eq.~\eqref{eq:compress-gr}, \({\left(\chi_{T}^{c}\right)}^{-1}\) is calculated.
    \item After both routes have been obtained, the squared difference of both quantities is calculated, i.e., \({\left[ {\left(\chi_{T}^{c} \left(\alpha, \beta\right) \right)}^{-1} - {\left(\chi_{T}^{v} \left(\alpha, \beta\right) \right)}^{-1} \right]}^2 \).
\end{enumerate}
 An schematic representation of these steps is shown in Fig.~\ref{fig:black-box-function}, where the different steps are shown in each square of the figure. The black-box function is comprised only on the middle part of the procedure, with the closure relation and the three parameter being the input, and the squared difference of the isothermal compressibilities is the output of the function.

Given that the function is not differentiable, DE methods were used to find the best set of parameters.
The general procedure is as follows:
\begin{enumerate}
    \item An initial search space is selected. In this work, the search space is chosen for each parameter: $A \in \left[-50, 0\right]$, $B \in \left[0, 50\right]$, $C \in \left[0, 50\right]$. An initial set of parameters is chosen randomly from within the search space.
    \item Using the DE global optimization method~\cite{wangDifferentialEvolutionBased2014}, a search is performed for all global minima. Note that the optimization procedure is constrained to the bounds of the search space.
    \item Only the parameters that evaluate to \(f(\bm{\alpha}^{\prime}) \leq 0.5\) are kept for the next step.
    \item Using a local optimization method, the SBPLX method~\cite{rowan1990functional,nlopt}, the parameters found by the previous two steps are taken as input for the function, \(\bm{\alpha}^{\prime}\). Now, the search bounds are restricted to $\bm{\alpha}^{\prime} \pm 1.5$ to force the optimizer to look for a better solution within the interval.
    \item The optimization procedure is stopped when the new set of parameters evaluate to \(f(\bm{\alpha}^{*}) \leq \num{1e-12}\).
\end{enumerate}
Thus, an optimal solution is found, $\bm{\alpha}^{*}$. These values are the ones that yield the same value
for the compressibility regardless the route taken. To obtain accurate estimations of the optimal solutions and the compressibility
values obtained, 15 independent solutions were determined using different random seeds and initial values.
The solutions were required to have positive isothermal compressibility values, and should also give
the same result regardless the route. Then, to compute the uncertainties
the bias-corrected bootstrap~\cite{efron1994introduction} was employed. To build 95\% confidence intervals,
$\num{e6}$ bootstrap samples were used.

\subsection{Numerical procedure}
In this work, two monocomponent fluids are considered. The first, the hard-sphere fluid,
described by the intermolecular potential,
\begin{equation}
    u_{\text{HS}} = \begin{dcases}
    \infty & r < \sigma \\
    0 & r \geq \sigma
    \end{dcases}
    \; ,
    \label{eq:hard-sphere}
\end{equation}
with $\sigma$ being the hard-core diameter.

The square well fluid was considered as well,
\begin{equation}
    u_{\text{SW}} = \begin{dcases}
    \infty & r < \sigma \\
    -\epsilon & \sigma \leq r \leq \lambda \sigma \\
    0 & r >\lambda \sigma
    \end{dcases}
    \; ,
    \label{eq:square-well}
\end{equation}
where $\lambda$ is the interaction range parameter and $\epsilon$ the well depth. For illustration purposes, in this work, only the value of $\lambda=1.5$ was considered.

For the solution of the OZ equation, an in-house code written in Julia~\cite{julia2017,ozjl} that implements a simple Picard scheme is used, which includes the five-point modification of Ng~\cite{ngHypernettedChainSolutions1974}.
A grid of 8192 equally spaced values was used to discretize the correlation functions. 
The numerical tolerance for the residual of the direct correlation functions during successive iterations
was set to $\num{e-5}.$ To reach
faster convergence, the solution to the OZ equation was started at a low density and using a grid
of 35 different values, the target density was reached, solving for each value the OZ equation and using
the result as an input for the next value. The integration was cut off at a value of $r_{c}=8 \sigma$.

For the compressibility route, the isothermal compressibility was determined using Eq.~\eqref{eq:compress-gr}, where the integral was computed using Romberg's method~\cite{press2007numerical}. For the virial route, the pressure was computed using the inverse of Eq.~\eqref{eq:pressure-gr}. A central difference scheme using a five-stencil rule~\cite{press2007numerical} was used to compute the isothermal compressibility, using different values for the density in order to obtain the numerical derivative. For each density value, the OZ equation is solved and the pressure is computed.
In general, the pressure is computed through the virial equation, which for the three-dimensional hard-sphere fluid~\cite{mcquarrieStatisticalMechanics2000} reads,
\begin{equation}
    P_{\text{HS}} = 1 + \frac{2 \pi}{3} \rho \sigma^{3} \, g(\sigma^{+}) \; .
    \label{eq:virial-hs}
\end{equation}
Similarly, for the square-well fluid, the pressure~\cite{mcquarrieStatisticalMechanics2000} reads,
\begin{equation}
    P_{\text{SW}} = 1 + \frac{2 \pi}{3} \rho \sigma^{3}
    \left(
    g(\sigma^{+}) + \lambda^{3} \left[ g(\lambda \sigma^{+}) 
    - g(\lambda \sigma^{-}) \right]
    \right) \; .
    \label{eq:virial-sw}
\end{equation}
In both expressions, the value $g(\sigma^{+})$ is the value of the radial distribution at contact,
as computed from the right side. Equivalently, the values $g(\lambda \sigma^{+})$ and 
$g(\lambda \sigma^{-})$ are the values of the RDF at the breakpoints at both sides of the potential range of the square-well fluid. In this work, all these quantities were obtained from the integral equation solutions
without extrapolation due to the grid points being sufficiently small such that the values at contact
could be obtained within the discretization of the correlation functions.

Furthermore, for certain closure relations, a different expression of the virial expression was
needed. For the Mean Spherical approximation, the expression due
to~\citeauthor{smithMeanSphericalApproximation1977}~\cite{smithMeanSphericalApproximation1977} was used.
This expression reads,
\begin{multline}
    P_{\text{MSA}} = 1 + \frac{2 \pi}{3} \rho \sigma^{3} g(\sigma^{+}) + \\*
    + \frac{\pi}{3} \beta \epsilon \rho \sigma^{3}
    \lambda^{3} \left[ g(\lambda \sigma^{+}) 
    - g(\lambda \sigma^{-}) \right]
    \; .
    \label{eq:msa-sw}
\end{multline}
For the Zerah-Hansen closure relation, the expression due to~\citeauthor{bergenholtzHMSAIntegralEquation1996}~\cite{bergenholtzHMSAIntegralEquation1996}
was used. This expression takes the following functional form,
\begin{multline}
    P_{\text{HMSA}} = 1 + \frac{2 \pi}{3} \rho \sigma^{3} g(\sigma^{+})
    - \frac{2 \pi}{3} \rho {(\lambda \sigma)}^{3} \times \\*
    \times \left(
    \beta \epsilon + \frac{1}{f(\lambda \sigma)}
    \left[ g(\lambda \sigma^{-}) 
    - g(\lambda \sigma^{+}) - \beta \epsilon \right]
    \right)
    \; .
    \label{eq:hmsa-sw}
\end{multline}
In this expression, the function $f$ is the mixing function of the Zerah-Hansen closure relation~\cite{zerahSelfConsistentIntegral1986},
taken to be 
$f(r)=1-e^{-ar},$ with $a$ being the free parameter to fix by enforcing some kind of thermodynamic
consistency. In this work, the Zerah-Hansen closure relation was solved using 
the Brent bracketing method~\cite{press2007numerical} to enforce compressibility consistency
and finding a suitable parameter for the mixing function.

\subsection{Computer simulations}
Computer simulations were used to compare the results obtained from integral equations theory. In particular,
the equation of state of the square-well fluid was obtained through event-driven molecular dynamics (EDMD)
using the DynamO~\cite{bannermanDynamOFreeCal2011} software. The simulations were performed for
systems with $N=16384$ particles, for a wide interval of densities $\rho \in [0.1, 0.8]$, in steps of $0.1.$
A single isotherm $T^{*}=k_{B} T / \epsilon = 1.5$ was studied, which is above the critical point for $\lambda=1.5.$~\cite{scholl-paschingerVaporliquidEquilibriumCritical2005b}
All systems were initialized using a face-centered cubic configuration. A Andersen thermostat~\cite{andersenMolecularDynamicsSimulations1980}
is used to thermalize the system in the $NVT$ ensemble to the specific reduced temperature.
The thermalization procedure consists of $\num{e8}$ collisions for each system. Then, the thermostat is removed
and the $NVE$ ensemble is sampled for an additional $\num{5e8}$ collisions. From these simulations,
the pressure is calculated through a time average over the change in momentum of
every collision~\cite{allen2017computer}. For each density value, 21 independent simulations were
done using different random seeds and initial values for the velocities of the particles.
To compute the statistical uncertainties, the bias-corrected bootstrap~\cite{efron1994introduction}
was employed. To build 95\% confidence intervals, $\num{e6}$ bootstrap samples were used.

%% file: files/results.tex
\section{Results and Discussion}\label{sec:results}

\subsection{Hard-sphere fluid}
\begin{figure}
    \centering
    \includegraphics[width=\linewidth]{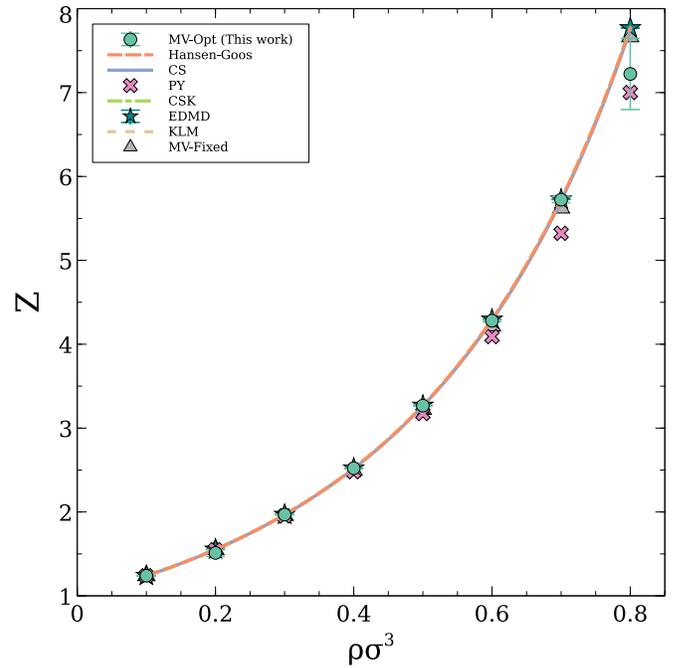}
     \caption{Compressibility factor of the hard-sphere fluid as a function of the reduced particle density, \(\rho \sigma^{3} .\) Only the results from event-driven molecular dynamics (EDMD) simulations and optimized modified Verlet closure have error bars (see section~\ref{sec:methods}). The comparison is done between the following different closure relations: the optimized, three-parameter, modified Verlet closure (MV-Opt)~\cite{verletIntegralEquationsClassical1981}; the Percus-Yevick (PY) closure relation~\cite{percusyevick1958}; the two-parameter modified Verlet closure relation (MV-Fixed)~\cite{verletIntegralEquationsClassical1980}. The rest of the comparison is made with the following theoretical equations of state, which basically predict the same compressibility factor within this density window: Hansen-Goos (HG)~\cite{hansen-goosAccuratePredictionHardsphere2016}; Kolafa-Lab{\'i}k-Malijevsk{\'y} (KLM)~\cite{kolafaAccurateEquationState2004}; Carnahan-Starling (CS)~\cite{carnahanEquationStateNonattracting1969}; and the Carnahan-Starling-Kolafa (CSK)~\cite{kolafaAccurateEquationState2004}.}
    \label{fig:hs-zfactor}
\end{figure}

%
\begin{table*}
\caption{Compressibility factor $Z=\beta P / \rho$ of the hard-sphere fluid for different values of the reduced number density, $\rho \sigma^{3} .$ The closure relation approximations used are the same as in Fig.~\ref{fig:hs-zfactor} and Fig.~\ref{fig:hs-relerror}. The number inside the parenthesis indicates the uncertainty of the last digits (see section~\ref{sec:methods}). The last row indicates results for the compressibility factor obtained through computer simulations using EDMD, as reported in \citeauthor{pieprzykThermodynamicDynamicalProperties2019}~\cite{pieprzykThermodynamicDynamicalProperties2019}.}
\centering
\resizebox{\textwidth}{!}{%
\begin{tabular}{c|cccccccc}
                & \multicolumn{8}{c}{Density, $\rho \sigma^{3}$}                                                      \\
\textbf{Closure approximation}         & 0.1        & 0.2        & 0.3        & 0.4        & 0.5        & 0.6        & 0.7       & 0.8       \\ \hline
\begin{tabular}[c]{@{}c@{}}Optimized\\ Verlet (This work)\end{tabular}    & 1.239678(62) & 1.512(69) & 1.9682(79) & 2.5218(35) & 3.2678(76) & 4.280(11) & 5.723(36) & 7.22(42) \\
Percus-Yevick   & 1.239329    & 1.549965    & 1.953768    & 2.480814    & 3.173219    & 4.091178    & 5.322839   & 7.000663   \\
MV-Fixed & 1.239365    & 1.550652   & 1.957950   & 2.496940   & 3.221974    & 4.218188    & 5.622829   & 7.664554   \\ \hline
\textbf{Equations of State}     & \multicolumn{8}{c}{} \\
Carnahan-Starling~\cite{carnahanEquationStateNonattracting1969} & 1.239666    & 1.553165   & 1.966711   & 2.518002   & 3.262431    & 4.283421    & 5.710209   & 7.749693 \\
Carnahan-Starling-Kolafa~\cite{kolafaAccurateEquationState2004} & 1.239717    & 1.553587   & 1.968191   & 2.521604   & 3.269514    & 4.295329    & 5.727445   & 7.769947 \\
Kolafa-Lab{\'i}k-Malijevsk{\'y}~\cite{kolafaAccurateEquationState2004} & 1.239720    & 1.553608   & 1.968229   & 2.521614   & 3.269392    & 4.294971    & 5.726998   & 7.770104 \\
Hansen-Goos~\cite{hansen-goosAccuratePredictionHardsphere2016} & 1.239720    & 1.553608   & 1.968227   & 2.521606   & 3.269358    & 4.294852    & 5.726655   & 7.769712 \\ \hline
\textbf{Simulation Methods}     & \multicolumn{8}{c}{}                                                                                \\
EDMD~\cite{pieprzykThermodynamicDynamicalProperties2019}            & 1.2397199(12) & 1.5536043(45) & 1.968231(11) & 2.521620(13) & 3.269404(11) & 4.294977(18) & 5.726960(29) & 7.770090(34) 
\end{tabular}%
}
\label{tab:hs-approx}
\end{table*}

In Table~\ref{tab:hs-approx} we compare the results for the compressibility factor
obtained for the hard-sphere fluid using several closure relations, as well as
theoretical equations of state. The EDMD simulations of \citeauthor{pieprzykThermodynamicDynamicalProperties2019}~\cite{pieprzykThermodynamicDynamicalProperties2019}
are taken as the ground truth for comparing the relative error in the estimation of the
compressibility factor. These results are highly accurate and were obtained through
extensive computer simulations.
All closure relations are in good agreement with computer simulations, up to the reduced density of $\rho \sigma^{3} = 0.6.$ For larger densities, the estimations provided by the mean-field equations
of state and the two-parameter MV closure are much better in terms of accuracy, when compared against computer simulations. Furthermore, it is
important to see that most of the results reported by Verlet~\cite{verletIntegralEquationsClassical1981}
are reproduced here, namely, the fact that the three-parameter MV closure relation has
good agreement to other equations of state and computer simulations, up to the
highest values of densities. In particular, this can be better observed in Fig.~\ref{fig:hs-zfactor}, where the compressibility factor for the highest density
reported start to slightly deviate from the other equations of state, which basically overlap each other, and the computer
simulation results.

\begin{figure}
    \centering
    \includegraphics[width=\linewidth]{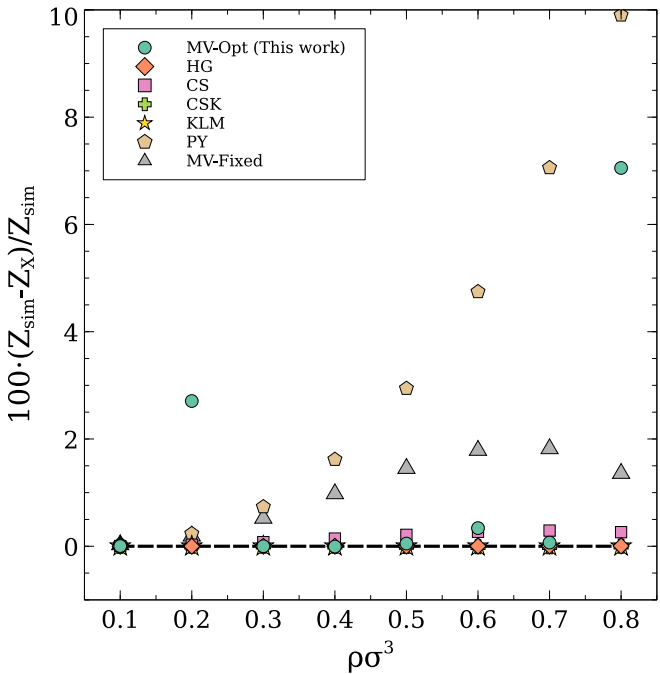}
    \caption{Relative percentage deviation \(100 \cdot \left( Z_{\text{sim}} - Z_{\text{X}}\right) / Z_{\text{sim}}\) as a function of the reduced particle density, $\rho \sigma^{3}$. $Z_{\text{sim}}$ is the compressibility factor obtained from EDMD simulations, as reported in \citeauthor{pieprzykThermodynamicDynamicalProperties2019}~\cite{pieprzykThermodynamicDynamicalProperties2019} for the hard-sphere fluid; see section~\ref{sec:methods}. $Z_{\text{X}}$ is the compressibility factor obtained from different closure relations, where the symbols have the same meaning as in Fig.~\ref{fig:hs-zfactor}.}
    \label{fig:hs-relerror}
\end{figure}

A clearer picture of the relative error is seen in Fig.~\ref{fig:hs-relerror}, where the
relative error between approximations and computer simulations is shown. In that figure, it is observed that the three-parameter MV closure relation
gives better results than the two-parameter MV closure relation for low to medium density
values. However, this changes at higher densities, where the two-parameter version is
clearly better. In the same way as Verlet reported his results, the three-parameter version
underestimates the compressibility factor at high densities. Although not shown here, it might
be possible that this behavior is due to the virial coefficients being underestimated as well,
given a poor prediction of the compressibility factor.
It is also interesting to note that the results obtained with the three-parameter version
of the MV closure relation gives good results when compared to the exact equation of state,
at least, from low to intermediate density values. When compared, the relative error is
quite small and gives a good estimate overall of the equation of state. This might seem that
the best closure relation overall is the three-parameter version of the MV for low to
intermediate density values and that it could be supplemented with the two-parameter version
for higher density values. In fact, if a fourth parameter were introduced to serve as
the mixing between the two-parameter and the three-parameter versions, the optimization
would remain the same and there would be almost no change in the computation time
and an estimation of the compressibility factor.

Most closure relations that enforce thermodynamic consistency do not have a unique set of
parameters that solve the optimization problem~\cite{tsedneeClosureOrnsteinZernikeEquation2019}.
A solution to this problem would be to increase the number of independent runs and
search for the variation in the parameters and see if within the range of the latter it is possible to find, in average,
the same local minima. This might be possible, but an important remark
is that at higher densities, the solution of the OZ equation and the search for a local
minima increases in difficulty. This can be seen in particular in the error bars from the compressibility factor of the hard-sphere fluid, Fig.~\ref{fig:hs-zfactor}. At the highest density explored, $\rho \sigma^{3} = 0.8$, the error is the highest from all the other density values explored, for the specific case of the three-parameter version of the Verlet closure. This points out the fact that at the density $\rho \sigma^{3} = 0.8$ the minima found were not accurate enough, due to poor performance of the optimization procedure, having found several minima with low accuracy and high variance. DE methods are suitable for highly nonlinear functions, but, 
in general, they do not always converge to global minimum~\cite{knoblochClassicDifferentialEvolution2017}.
If this is the case, then the local optimizer will not find a suitable minimizer to
the function. A possible solution to the problem is to incorporate even more physical
information into the loss function, Eq.~\eqref{eq:optimiziation-chi}. For instance,
if it could incorporate a simple regularization term~\cite{hastie2009elements}, it might
be possible to increase the number of successful solution for the optimization problem.

\subsection{Square-well fluid}
\begin{figure}
    \centering
    \includegraphics[width=\linewidth]{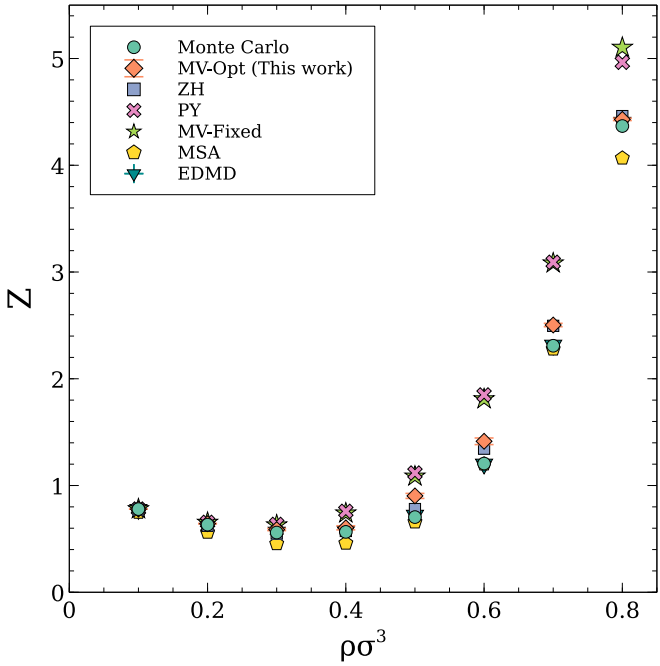}
    \caption{Compressibility factor of the square-well fluid for an attraction parameter of $\lambda=1.5$, at a reduced temperature of $T^{*}=1.5$, as a function of the reduced particle density, \(\rho \sigma^{3} .\) Only the results from event-driven molecular dynamics (EDMD) simulations and optimized modified Verlet closure have error bars (see section~\ref{sec:methods}). The results of the following closure relations are also displayed: the optimized, three-parameter, modified Verlet closure (MV-Opt)~\cite{verletIntegralEquationsClassical1981}; the Zerah-Hansen (ZH)~\cite{zerahSelfConsistentIntegral1986}; the Percus-Yevick (PY)~\cite{percusyevick1958}; the two-parameter modified Verlet closure (MV-Fixed)~\cite{verletIntegralEquationsClassical1980}; and the Mean Spherical Approximation (MSA)~\cite{smithMeanSphericalApproximation1977}. The data here shown are also summarized in Table~\ref{tab:sw-approx}.}
    \label{fig:sw-zfactor}
\end{figure}

To test the ability of estimating the compressibility factor with the three-parameter
MV closure relation, we have also studied the square-well fluid for a single isotherm and a single
interaction range, namely, $T^{*}=1.5$ and $\lambda=1.5$, respectively. The results obtained with other closure relations, as well as the computer
simulations are displayed in Fig. \ref{fig:sw-zfactor} and summarized in Table~\ref{tab:sw-approx}. For this intermolecular potential, the situation
is quite different when compared to the hard-sphere fluid. As we mentioned previously, to the best of our
knowledge, this is the first attempt of using the three-parameter version of the MV closure
relation with the square-well fluid.

\begin{figure}
    \centering
    \includegraphics[width=\linewidth]{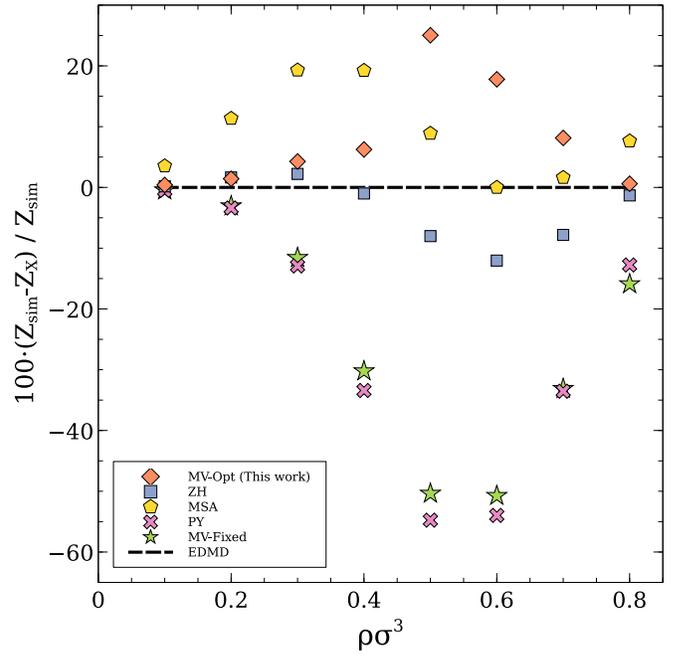}
    \caption{Relative percentage deviation \(100 \cdot \left( Z_{\text{sim}} - Z_{\text{X}}\right) / Z_{\text{sim}}\) as a function of the reduced particle density, \(\rho \sigma^{3}.\) \(Z_{\text{sim}}\) is the compressibility factor obtained from EDMD simulations for a square-well fluid with an attraction parameter of \(\lambda=1.5\) at reduced temperature \(T^{*}=1.5\); see section~\ref{sec:methods}. \(Z_{\text{X}}\) is the compressibility factor obtained from different closure relations, where the symbols have the same meaning as in Fig.~\ref{fig:sw-zfactor}.}
    \label{fig:sw-relerror}
\end{figure}

From Fig.~\ref{fig:sw-zfactor}, it is evident that at intermediate densities, $\rho \sigma^{3} \in [0.3, 0.6]$, all closure relations give different results. The relative error reported in Fig.~\ref{fig:sw-relerror} clearly illustrates this situation. A similar result
was reported by~\citeauthor{bergenholtzHMSAIntegralEquation1996}~\cite{bergenholtzHMSAIntegralEquation1996},
where the use of the Zerah-Hansen closure relation gives good results when compared to 
accurate results from computer simulations, mostly at a density region that does not include
high densities. The same results were also found in this work, i.e., the Zerah-Hansen closure 
relation becomes more accurate than the other closure relations explored. However, when
compared with the three-parameter MV closure relation, it can be seen that both 
approximations give similar results, with the difference that the Zerah-Hansen closure underestimates
the compressibility factor, while the three-parameter MV closure relation overestimates it.
This indicates that if the results of both closure relations were, in some way, combined
or mixed, both closures could complement each other and might give better results. With the idea of mixing the three-parameter MV closure relation with the Zerah-Hansen in the same optimization problem, it could be possible to provide
a better estimation of the compressibility factor, with small changes to the optimization
procedure. However, this case is neither explored nor reported in this contribution.

For the lowest density studied, $\rho \sigma^{3}=0.1,$ almost all closure relations give the same value and agree well with
computer simulation results. However, for the highest density case, $\rho \sigma^{3}=0.9,$ this is not longer the case.
In contrast to the hard-sphere fluid, it seems that higher densities are not really
a problem for the square-well fluid. Only the two-parameter version of the MV and the
Percus-Yevick give poor results when compared with the computer simulation results,
as seen in Fig.~\ref{fig:sw-zfactor}. On the other hand, the three-parameter version gives
excellent results, quite similar to those obtained by the Zerah-Hansen closure relation.
The Mean Spherical Approximation is also a good closure relation for the square-well fluid, 
and this has been previously reported~\cite{smithMeanSphericalApproximation1977,smithCorrectedIntegralEquations1978}.

%

%
\begin{table*}
\caption{Compressibility factor $Z=\beta P / \rho$ for different values of the reduced number density, $\rho \sigma^{3},$ using several closure approximations for the square-well fluid with \(\lambda=1.5\). The closure relation approximations used are the same as the ones reported in Fig.~\ref{fig:sw-zfactor} and Fig.~\ref{fig:sw-relerror}. The number inside the parenthesis indicates the uncertainty of the last digits (see section~\ref{sec:methods}). The last two rows correspond to the compressibility factor obtained through both EDMD and Monte Carlo computer simulation methods. MC data was taken from Ref. \cite{largoPairCorrelationFunction2005}.}
\centering
\resizebox{\textwidth}{!}{%
\begin{tabular}{c|dddddddd}
                & \multicolumn{8}{c}{Density, $\rho \sigma^{3}$}                                                      \\
\textbf{Closure Approximation}         & 0.1        & 0.2        & 0.3        & 0.4        & 0.5        & 0.6        & 0.7       & 0.8       \\ \hline
\begin{tabular}[c]{@{}c@{}}Optimized\\ Verlet (This work)\end{tabular}    & 0.7789(47) & 0.642(56) & 0.58(19) & 0.60(38) & 0.91(5) & 1.41(6) & 2.50(3) & 4.42(2) \\
Zerah-Hansen    & 0.77446    & 0.62316    & 0.55054    & 0.57412    & 0.77977    & 1.34523    & 2.49603   & 4.46077   \\
Percus-Yevick   & 0.77952    & 0.65493    & 0.63587    & 0.75835    & 1.11718    & 1.84873    & 3.09157   & 4.96477   \\
\begin{tabular}[c]{@{}c@{}}Mean Spherical\\ Approximation\end{tabular} & 0.74848    & 0.56162   & 0.45445  & 0.45907  & 0.65774 & 1.20081 & 2.27751 & 4.06664 \\
Modified Verlet & 0.77936    & 0.653043   & 0.628289   & 0.740228   & 1.08556    & 1.80991    & 3.08212   & 5.10293   \\ \hline
\textbf{Simulation methods}     & \multicolumn{8}{c}{}                                                                                \\
EDMD            & 0.7758(36) & 0.6335(28) & 0.5631(41) & 0.5684(39) & 0.7219(79) & 1.2007(90) & 2.315(13) & 4.403(23) \\
Monte Carlo~\cite{largoPairCorrelationFunction2005}     & 0.77902    & 0.63332    & 0.559392   & 0.565727   & 0.704298   & 1.20577    & 2.30921   & 4.36758  
\end{tabular}%
}
\label{tab:sw-approx}
\end{table*}

The reduced densities $\rho \sigma^{3}=0.5,0.6$ will now be investigated.
We begin with the value of $\rho \sigma^{3}=0.5$. No closure relation used in this work
can estimate the compressibility factor for this density accurately. Looking
at the relative error in Fig.~\ref{fig:sw-relerror}, it can be seen that the only closure
relations that come close to a good approximation are the Mean Spherical Approximation and
the Zerah-Hansen, with the Zerah-Hansen being the closest with a relative error of 
approximately -8\%, which means that it overestimates the compressibility factor.
The three-parameter MV approximation deviates quite heavily, surpassing
20\% of the relative error, which is quite high. Furthermore, the two-parameter version and
the Percus-Yevick equation deviate for more than 50\%, i.e., both closures are clearly inaccurate and not useful at all.
For the case of the $\rho \sigma^{3}=0.6$, the situation is similar, but it is now observed
that the Mean Spherical Approximation can estimate the compressibility value with high
accuracy. However, the Zerah-Hansen and the three-parameter MV still deviate quite strongly,
with the Zerah-Hansen being more accurate. It is not really clear why this is the case,
as it would seem by observing the adjacent density values of $\rho \sigma^{3}=0.4$ and
$\rho \sigma^{3}=0.7$, that the three closure relations oscillate between their estimations.
For the reduced density $\rho \sigma^{3}=0.4$, the Zerah-Hansen is better; for
 $\rho \sigma^{3}=0.5$ it has already been shown that neither closure relation is
good enough; for $\rho \sigma^{3}=0.6$, the Mean Spherical Approximation is better
and, finally, for $\rho \sigma^{3}=0.7$ is estimated almost equally by the three,
with the Mean Spherical Approximation the best.

The Percus-Yevick and the two-parameter MV closure relations are the least accurate of all closure relations explored.
It is known that the Percus-Yevick does not give good estimates for the compressibility factor of the square-well
fluid~\cite{barkerPerturbationTheoryEquation1967,smithPercusYevickEquation1974a}.
This fact is clearly shown in the results of this work, where the Percus-Yevick is the worst
closure relation in terms of accurately estimating the compressibility factor. At low densities,
the results are quite similar to other closure relations, all estimate the compressibility factor quite well.
At intermediate densities, the Percus-Yevick approximation deviates greatly from computer simulation results and
other closure relations.
For the case of the two-parameter MV closure relation, the situation is quite similar. In contrast, this closure relation was recently used to study the thermodynamic, static and dynamic
properties of competing interaction fluids~\cite{perdomo-perezThermodynamicsStaticProperties2022}.
For an interaction parameter of $\lambda=1.15$, \citeauthor{perdomo-perezThermodynamicsStaticProperties2022} found that the two-parameter
version was an accurate closure relation for the square-well fluid, which was one of the
systems studied in that work. Yet, the comparison cannot be done directly with this work,
due to the attraction well being smaller, and somewhat closer to the hard-sphere fluid.
In that case, the two-parameter MV closure relation could give good results for high temperatures (above the binodal), as shown in the results obtained in this work for the hard-sphere fluid. Results for square-well fluids with a shorter interaction range will be reported elsewhere.